# CHARGE ORDERING AND PHASE COMPETITION IN THE LAYERED PEROVSKITE LaSr$_2$Mn$_2$O$_7$

D. N Argyriou[1†], H. N. Bordallo[1‡], B. J. Campbell[2], A. K. Cheetham[2]. D. E. Cox[3], J. S. Gardner[4¶], K. Hanif[5], A. dos Santos[2], G. F. Strouse[5].

[1]*Los Alamos Neutron Science Center, Los Alamos National Laboratory*
[2]*Materials Research Laboratory, University of California, Santa Barbara, CA, 93106.*
[3]*Physics Department, Brookhaven National Laboratory, Upton, NY 11973.*
[4]*MST-10, Los Alamos National Laboratory, Los Alamos, NM, 80434.*
[5]*Department of Chemistry, University of California, Santa Barbara, CA 93106.*

Charge-lattice fluctuations are observed in the layered perovskite manganite LaSr$_2$Mn$_2$O$_7$ by Raman spectroscopy as high as 340 K and with decreasing temperature they become static and form a charge ordered (CO) phase below T$_{CO}$=210 K. In the static regime, superlattice reflections are observed through neutron and x-ray diffraction with a propagation vector (h+1/4,k-1/4,l). Crystallographic analysis of the CO state demonstrates that the degree of charge and orbital ordering in this manganite is weaker than the charge ordering in three dimensional perovskite manganites. A T$_N$=170K a type-A antiferromagnetism (AF) develops and competes with the charge ordering, that eventually melts below T*=100K. High resolution diffraction measurements suggest that that CO- and AF-states do not coincide within the same region in the material but rather co-exist as separate phases. The transition to type-A antiferromagnetism at lower temperatures is characterized by the competition between these two phases.

## I. INTRODUCTION

The competition between charge, lattice and spin degrees of freedom can be delicately balanced to form materials where electrons localize on alternate transition metal sites on a lattice. This so-called *charge ordered* (CO) lattice has been observed for a number of transition metal perovskites and is fundamental in the understanding of the physical properties of many of these materials. For example, the formation of dynamic charge ordering fluctuations has recently been reported in superconducting La$_{1.85}$Sr$_{0.15}$CuO$_4$,[1] and YBa$_2$Cu$_3$O$_{7-x}$[2] while in Pr$_{0.7}$Ca$_{0.3}$MnO$_3$ the formation of a ferromagnetic metallic phase by irradiation with X-rays arises from the melting of a charged ordered lattice.[3] The fundamental physics governing these phenomena are applicable to a wide range of materials such as linear chain systems, conducting organic materials and magnetic semiconductors.[4,5]

The observation of colossal negative magnetoresistance (CMR) in the layered manganites La$_{2-2x}$Sr$_{1+2x}$Mn$_2$O$_7$[6] has provided the opportunity to study the strong interplay among charge, spin and lattice in reduced dimensions, and also to explore novel phenomena that are not found in the 3D perovskite manganites. For example, the natural stacking of (La,Sr)MnO$_3$ perovskite bi-layers separated by (La,Sr)O blocking layers forms the framework of tunneling structures and spin valves in a single chemical phase, as opposed to the larger scale structures obtained by thin film deposition.[7,8]

Recently, CO has been reported in layered LaSr$_2$Mn$_2$O$_7$ (x=0.5). However, unlike the 3D perovskite compounds, the CO state is stable only over a limited temperature region (~200-100K).[9] This suggests that the lower dimensionality of this material effects directly the competition between charge, lattice and spin degrees of freedom. Charge and orbital ordering in the manganite perovskites brings into focus the overall issue of the influence of electron-phonon coupling on the transport and magnetic properties of these materials.

In this paper we report neutron, X-ray and resonant Raman scattering measurements on LaSr$_2$Mn$_2$O$_7$. We find that charge-lattice fluctuations arising from perturbations in the local crystal structure from hopping e$_g$ carriers are evident in resonant Raman measurements as high as 340K. With decreasing temperature these fluctuations become static below T$_{CO}$=210 K and superlattice reflections indicative of charge ordering are observed using both neutron and x-ray diffraction. Analysis of single crystal neutron diffraction data shows that the structure of LaSr$_2$Mn$_2$O$_7$ is characterized by partial charge localization, as measured from the degree of structural distortion around nominally Mn$^{3+}$ and Mn$^{4+}$ sites. With decreasing temperature the development of the CO phase is disrupted by the onset of a type-A antiferromagnetic (AF) ordering of Mn-spins at $T_N$=170 K. As the CO lattice melts and disappears below T*=100K, charge-lattice fluctuations reappear and remain to low temperatures. The competition between the charge ordered and type-A AF phase is highlighted by the broadening and eventual split of diffraction profiles. This behavior suggests that charge ordering and antiferromagnetism do not coincide in these materials but rather co-exist as separate phases. At low temperature we find evidence for only a type-A antiferromagnetic insulating phase.

---

[†] Present address: Materials Science Division, Argonne National Laboratory, Argonne, IL, 60439.

[‡] Present address: Intense Pulsed Neutron Source, Argonne National Laboratory, Argonne, IL, 60439.

[¶] NPMR, Chalk River Laboratories, Bld 459, Stn. 18, Chalk River, ON, KOJ 1JO, Canada.

## II. EXPERIMENTAL

Neutron diffraction data from a single crystal of $LaSr_2Mn_2O_7$ (x=0.5) were measured as a function of temperature (300-20K) using the single crystal diffractometer (SCD) at the Los Alamos Neutron Science Center, Los Alamos National Laboratory. The crystal was cleaved from a boule prepared by the optically floating zone method. Single crystals from this boule have also been characterized with resistivity and Inductively Coupled Plasma (ICP) spectroscopy measurements. The resistivity measurements in the *ab*-plane showed a broad peak between 210-150 K associated with charge ordering (see fig. 1(a)), and are in agreement with the measurements of Kimura *et al*.[9] ICP measurements gave a composition of x=0.50(2) for the single crystal sample.

A powder sample of $LaSr_2Mn_2O_7$ was also prepared from sol-gel precursors. The samples were prepared by dissolving stoichiometric amounts of $La(OH)_3$, MnO, and $SrCO_3$ in glacial acetic acid and stirring for several hours at about 80 C. After boiling off all the acid, the resulting pink gel was dried at 400C for 4 hours, annealed at 1400 °C for 24 hours, and quenched to room temperature, all in air. The powder sample was characterized by X-ray powder diffraction measured from 300-20 K on the high-resolution powder diffractometer X7A at the National Synchrotron Light Source, Brookhaven National Laboratory. For this experiment the sample was mounted in a 0.4 mm glass capillary and data was collected using a wavelength of 0.8 Å and a Ge(111)/Ge(220) monochromator/analyzer combination. From the line widths of the X-ray diffraction data we estimated the compositional fluctuations to be ±0.02 (i.e. x=0.50(2)).

The powder sample was also used in measuring resonant Raman scattering as a function of temperature (16-340K) using laser wavelengths of 2.41, 2.54 and 2.7 eV. Measurements were made in a back-scattering configuration using an $Ar^+$-ion laser excitation (3 mW incident power) coupled to a Notch filter/ 0.5M ARC single spectrograph (1800 lines mm-1/400 blaze) - Princeton Instrument 512x512 CCD array for detection. Temperature control at ±1.0K was provided by an APD closed cycle He-cryostat controlled by a SI9620-1 temperature controller. Our data confirmed that power densities over 5 mW results in irreversible photo-damage to the manganite sample, as observed previously in $LaMnO_3$ by Iliev *et al*.[10]

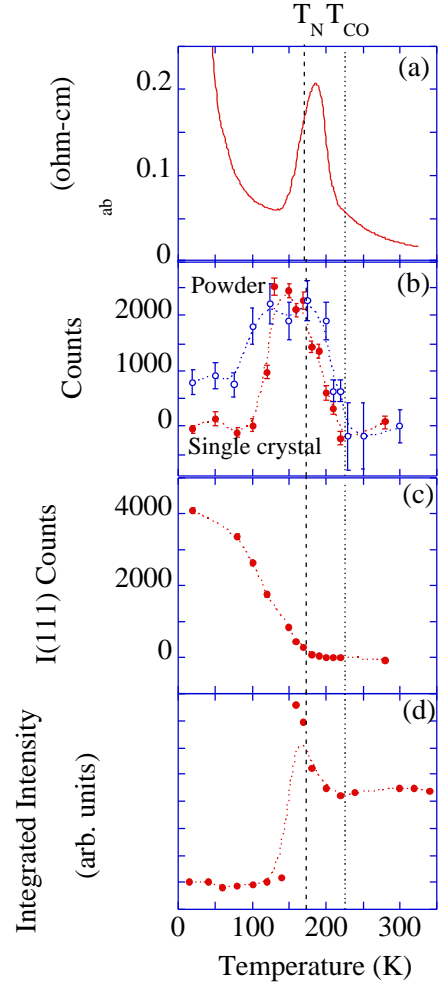

FIG. 1. (a) In-plane resistivity as a function of temperature from a $LaSr_2Mn_2O_7$ single crystal. (b) Temperature dependence of the (1.75,2.25,0) super lattice reflection measured from the single crystal sample using neutron diffraction (filled symbols) and the (0.25, 1.75,0) superlattice reflection measured from the powder sample using x-ray diffraction (open symbols). (c) Temperature dependence of the (111) type-A antiferromagnetic reflection measured from the single crystal sample using neutron diffraction. (d) Temperature dependence of the frequency of the Mn-O(3) bond bending mode ($\omega_3$) and the coupled resonant mode ($\omega_{3'}$), measured from the powder sample using Raman scattering with excitation wavelength of 514 nm. Doted lines are a guide to the eye.

## III. CHARGE ORDERING AND PHASE CO-EXISTENCE IN LaSr$_2$Mn$_2$O$_7$

Room temperature neutron and X-ray diffraction measurements for both single crystal and powder samples were consistent with the well-known I4/*mmm* crystal structure.[11] On cooling these samples below $T_{CO}$=210K weak superlattice reflections with a propagation vector **q**=($h$+1/4, $k$-1/4, $l$) were observed, in agreement with the measurements reported by Kimura *et al.*[9]. The intensities of the superlattice reflections track approximately the in-plane resistivity, reaching a maximum at ~150K with decreasing temperature and disappearing below T*=100K for the single crystal neutron measurements (see fig. 1(b)).[9] This is in sharp contrast to the perovskite La$_{1/2}$Ca$_{1/2}$MnO$_3$ where the CO state is stable to at least 5K.[12] For the powder sample we find a similar decrease in intensity at ~90 K but the superlattice reflections do not disappear completely and are still visible at 20K (see fig 1(b)). We attribute this to extrinsic effects such as pinning by grain boundaries or point defects. We also note that the widths of the superlattice reflections in both the single crystal and powder samples remained close to the instrumental resolution over the temperature range that they were observed, suggesting coherence lengths of 100-1000s of Angstroms. In addition to these superlattice reflections, magnetic reflections are observed below $T_N$=170K in our single crystal measurements, that are consistent with a type-A AF phase (see fig. 1(c)).

A detailed insight and a crystallographic description of the charge and orbital ordering in LaSr$_2$Mn$_2$O$_7$ is obtained by analyzing the single crystal neutron diffraction data measured at 160K. The additional superlattice reflections can be indexed on a commensurate orthorhombic super cell of $a_o$~ $2a_t$ ,$b_o$~$2\sqrt{2}a_t$  $c_o$=$c_t$ (where $a_t$ and $c_t$ are the parent tetragonal lattice constants) as shown in fig. 2. The observation of superlattice reflections by neutrons and X-rays clearly suggest that they arise from structural changes as opposed to magnetic ordering as suggested by Kubota *et al.*[13] In addition recent polarized neutron measurements from the same single crystal sample used in this work, clearly showed that the super lattice reflections reported here, exhibit no significant magnetic component.[14] The observed extinction conditions for the super structure reflections are consistent with space group B*bmm*. This reduced symmetry provides for two different Mn sites (Mn(1) and Mn(1´)), where the surrounding in-plane oxygens can relax in both the $a_o$ and $b_o$ directions (see fig. 2) compared to the parent I4/*mmm* structure. We note here that between $T_{CO}$ and T* only an *average* structure description of LaSr$_2$Mn$_2$O$_7$ can be obtained using the parent space group symmetry I4/*mmm*. The details of the charge ordered structure are contained entirely in the weaker super lattice reflections.

In the refinement of the structure, we found that displacements from the average structure along the $b_O$ and $c$ directions did not compute significant intensity for super lattice reflections while displacements along $a_O$ did. In the final model, only displacements along $a_O$ were refined for the Mn (1´) site (labeled at Mn$^{4+}$ in fig. 2), its surrounding oxygen atoms (O(3), O(3´), O(1´) and O(2´)) and the La,Sr atoms above and below the Mn(1´) site (La,Sr(1´) and La,Sr(2´)). The super structure was refined using only superlattice reflections as their low intensity (~0.7% of the parent Bragg reflection) do not contribute significantly to the overall $\chi^2$ of the refinement when all observed reflections are considered. The final refinement produced a weighted R factor, wR(F$^2$), of 20.1% for the refinement using superlattice reflections only, and 7.75% for all reflections. Refined crystallographic parameters for the charge ordered structure of LaSr$_2$Mn$_2$O$_7$ are given in table I.

Table I: Crystallographic parameters for the CO state of LaSr$_2$Mn$_2$O$_7$ at 160 K as measured using single crystal neutron diffraction measurements. For the average structure refinements, 900 reflections were used and a wR(F$^2$) of 7.8 % was obtained. For the super structure refinements, 1901 super structure reflections were measured, 1046 of which has I>0 and 68 had I>3$\sigma$. The lattice constants were found to be $a$=5.443(4) Å,$b$=10.194(4) Å,$c$=19.816(15) Å. The y coordinates were not refined. Temperature factors were constrained as follows U(Mn(1))=U(Mn(1´)), U(La,Sr(1))=U(La,Sr(1´))=U(La,Sr(2))=U(La,Sr(2´)), and for the oxygen atoms U(O(1))=U(O(1´)), U(O(2))=U(O(2´)), U(O(3))=U(O(3´)). Various Mn-O bond lengths computed from the structural parameters are also given below.

| | x | y | z | Uiso (Å$^2$) |
|---|---|---|---|---|
| Mn$^{3+}$(1) | 0 | 0 | 0.09696(4) | 0.0021(1) |
| Mn$^{4+}$(1´) | 0.5065(2) | 0.25 | 0.09696(4) | 0.0021(1) |
| La,Sr(1) | 0 | 0 | 0.5 | 0.0051(1) |
| La,Sr(1´) | 0.5218(5) | 0.25 | 0.5 | 0.0051(1) |
| La,Sr(2) | 0 | 0 | 0.31788(2) | 0.0051(1) |
| La,Sr(2´) | 0.5073(5) | 0.25 | 0.31788(2) | 0.0051(1) |
| O(1) | 0 | 0 | 0 | 0.0098(2) |
| O(1´) | 0.5164(7) | 0.25 | 0 | 0.0098(2) |
| O(2) | 0 | 0 | 0.19505(3) | 0.0099(4) |
| O(2´) | 0.5035(4) | 0.25 | 0.19505(3) | 0.0099(4) |
| O(3) | 0.2588(2) | 0.125 | 0.09495(2) | 0.0077(1) |
| O(3´) | 0.7510(2) | 0.125 | 0.09495(2) | 0.0077(1) |

| | | | |
|---|---|---|---|
| Mn$^{3+}$(1)-O(3´) | 1.962(1) Å | Mn$^{4+}$(1)-O(3´) | 1.906(1) Å |
| Mn$^{3+}$(1)-O(3) | 1.923(1) Å | Mn$^{4+}$(1)-O(3) | 1.918(1) Å |
| Mn$^{3+}$(1)-O(1) | 1.921(2) Å | Mn$^{4+}$(1)-O(1´) | 1.921(2) Å |
| Mn$^{3+}$(1)-O(2) | 1.944(2) Å | Mn$^{4+}$(1)-O(2´) | 1.944(2) Å |

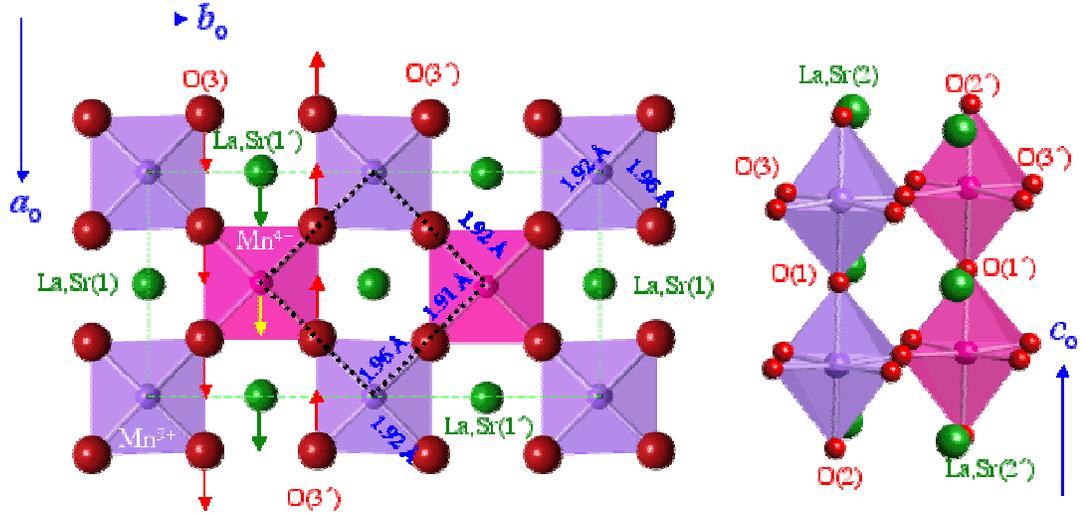

FIG. 2. The crystal structure of the CO state of $LaSr_2Mn_2O_7$. The black thick lines show the cell of the parent I4/*mmm* structure, while the thinner green lines shows the size of the $2a_t \times 2 \; 2a_t \times c_t$ super cell. Arrows show the displacement of atoms from the parent I4/*mmm* structure. Atoms labeled as (´) indicate equivalent sites in the I4/*mmm* setting. Various Mn-O bond lengths are shown.

Two distinct Mn sites are identified from the analysis of the single crystal neutron diffraction data (see figure 2), consistent with a structural distortion arising from the localization and ordering of $e_g$ carriers. The Mn(1) site exhibits two short in-plane Mn-O bonds of 1.923(1) Å and two long bonds of 1.962(1) Å, the latter corresponding to occupied $d_{3z^2-r^2}$ orbitals extended along the bond axis. The second site, Mn(1´), has two similar pairs of in-plane Mn-O bonds of 1.918(1)Å and 1.906(1)Å (see fig. 2). The apical oxygen distances for both Mn sites are, 1.921(2) Å for the Mn-O(1) and 1.944(2)Å for the Mn-O(2) bond. Differences in electron density between the two Mn sites are evident in the bond valence sums, $v_{ij}$, where we obtain values of 3.67 and 3.87 for the Mn(1) and Mn(1´) sites respectively. These values suggest that the ionic model of CO in $LaSr_2Mn_2O_7$, in which discreet $Mn^{3+}$ and $Mn^{4+}$ sites exist due to the localization of $e_g$ carriers, is oversimplified and a large degree of covelency involving Mn $d$ orbitals and O $p$ orbitals may exist.

The structural model of the charge ordered state in $LaSr_2Mn_2O_7$ presented is essentially the same as that proposed by Goodenough[15] and recently observed by Radaelli *et al.*[12] for the manganite perovskite $La_{1/2}Ca_{1/2}MnO_3$. However, a clear difference exists in bond valences of the Mn sites. For $La_{1/2}Ca_{1/2}MnO_3$, the nominally $Mn^{4+}$ and $Mn^{3+}$ sites have bond valance values of 3.9 and 3.5, respectively,[12] suggesting that the degree of charge ordering in $LaSr_2Mn_2O_7$ is smaller than that observed in the perovskite manganites. In addition, the partial *charge* ordering is also reflected in a partial *orbital* ordering; charge is shared unequally between $d_{x^2-y^2}$ and $d_{3z^2-r^2}$ orbitals. Typically, filled $Mn^{3+} d_{3z^2-r^2}$ orbitals produce a bond length of ~2.1 Å, larger than the value of 1.962 Å observed for $LaSr_2Mn_2O_7$, while empty $d_{x^2-y^2}$ orbitals have a bond length of 1.91 Å, smaller than the 1.923 Å observed in $LaSr_2Mn_2O_7$.[16] These values suggest that the former orbital is under-occupied while the latter is not completely empty.

At low temperatures the neutron diffraction measurements show that nuclear reflections are consistent with the well know I4/*mmm* crystal structure[11], while new magnetic reflections indicate a type-A antiferromagnetic ordering.[17] Here, Mn moments form ferromagnetic $MnO_2$ sheets that are antiferromagnetically coupled along the *c*-axis and within a perovskite bi-layer (see table II for crystal and magnetic parameters determined at 20K). That the orbital ordering observed in $LaSr_2Mn_2O_7$ and $La_{1/2}Ca_{1/2}MnO_3$ is essentially identical would suggest that a similar magnetic ordering would be observed in both compounds. As described by Goodenough[15] in the 1950's, the ordering of Mn $d_{3z^2-r^2}$ orbitals in $La_{1/2}Ca_{1/2}MnO_3$ gives rises to staggered ferromagnetic interactions between filled $Mn^{3+}$ *d*- and empty $Mn^{4+}$ *d*-orbitals and antiferromagnetic interactions between empty $Mn^{3+}$ *d*- and $Mn^{4+}$ *d*-orbitals (the so called CE magnetic structure). This is in stark contrast to the observation of ferromagnetic sheets in $LaSr_2Mn_2O_7$ and suggests that in the case of weak charge and orbital ordering, Goodenough's rules for ferromagnetic and antiferromagnetic $Mn^{3+}$-O-$Mn^{4+}$ coupling may not hold entirely.

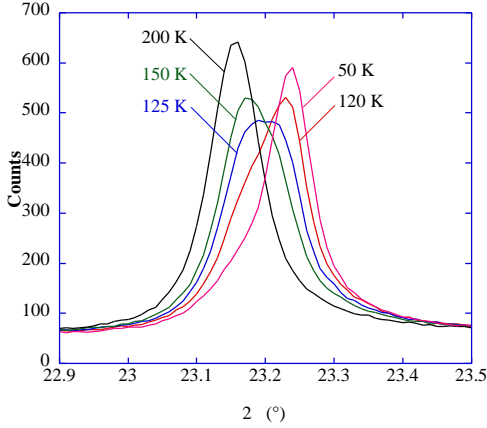

FIG. 3. The temperature dependence of the (0 0 10) peak profile measured using high-resolution x-ray diffraction.

Table II: Crystal and magnetic structure parameters for $LaSr_2Mn_2O_7$ at 20 K determined using single crystal neutron diffraction. At 20K no super lattice reflections were observed. The crystal structure was analyzed using the tetragonal I4/*mmm* crystal structure[11]. The magnetic structure was analyzed in term of type-A Mn spin arrangement; Mn spins lie ferromagnetically within $MnO_2$ along the *ab*-plane, while the coupling between sheets is antiferromagnetic.[17] In all 624 reflections were measured. Least squares refinement of the model produced a $wR(F^2)$ of 9 %. The lattice constants were found to be $a$=3.852(3) Å, $c$=19.76(2) Å. Various Mn-O bond lengths computed from the structural parameters are also given below.

| | x | y | z | Uiso (Å$^2$) |
|---|---|---|---|---|
| $Mn^{3+}(1)$ | 0 | 0 | 0.09701(16) | 0.005(3) |
| $\mu_{ab}=$ | 2.63(6) | | | |
| La,Sr(1) | 0 | 0 | 0.5 | 0.0012(4) |
| La,Sr(2) | 0 | 0 | 0.31819(8) | 0.0024(3) |
| O(1) | 0 | 0 | 0 | 0.0064(6) |
| O(2) | 0 | 0 | 0.19516(13) | 0.0065(5) |
| O(2´) | 0.5 | 0 | 0.09490(9) | 0.0047(3) |

| | |
|---|---|
| Mn(1)-O(1) | 1.916(4) Å |
| Mn(1)-O(2) | 1.939(5) Å |
| Mn(1)-O(3) | 1.9291(13) Å |

The resolution of this apparent controversy between the observation of CE-type charge and orbital ordering and type-A antiferromagnetic ordering lies in the competition of between the CO- and AF-phases. Below $T_{CO}$, that the (0 0 10) reflection at high temperatures is symmetrical, the only significant change being the continuous shift towards higher $2\theta$ as the *c*-axis lattice parameter contracts with decreasing temperature (Fig. 4). Below $T_N$ a high-angle shoulder develops, marking the formation of a new phase with a slightly lower value of the *c*-axis lattice parameter. The loss of intensity from the original peak makes it clear that the new magnetically ordered phase is forming at the expense of the CO phase formed at a higher temperature. By 125 K, there are nearly equal amounts of the two phases, and by 50 K the new phase has almost entirely consumed the first. This suggests that the type-A AF-phase and the CO-phases do not coincide in the material but rather co-exist as separate and distinct phases. We note that this phase co-existence separates the magnetic and charge-lattice degrees of freedom in the system and thus lifts the apparent controversy between CE-type charge ordering and type-A antiferromagnetism.

The development of the AF phase correlates with a decrease in the resistivity of the material below $T_{CO}$, and suggests that charge carriers in the AF-phase become delocalized compared to the CO-phase. Thus the co-existence here between the CO- and AF-phases also highlights the co-existence of two different electronic states. A similar picture has been proposed for the co-existence of a ferromagnetic metallic phase and a charge ordered phase in $Pr_{0.7}Ca_{0.3}MnO_3$.[3,18] Indeed the photo-induced insulator-metal transition observed in that material is somewhat analogous to the transition from charge ordering to an antiferromagnetic insulator in $LaSr_2Mn_2O_7$. However, the *electronic phase separation* in $Pr_{0.7}Ca_{0.3}MnO_3$ is somewhat different to what we find in $LaSr_2Mn_2O_7$ in that charge in both CO- and AF-phases is essentially localized and the separation occurs between a charge ordered state and an antiferromagnetic insulator.

We note here that although the orbital ordering we observe is consistent with the CE-antiferromagnetic structure, the observation of type-A antiferromagnetism is suggestive of a completely different orbital ordering. Mizokawa and Fujimori[19] point out that type-A antiferromagnetism is consistent with charge ordering were the $e_g$ charge resides in $Mn^{3+}$ $d_{x^2-y^2}$ orbitals in the *ab*-plane as opposed to $d_{3z^2-r^2}$ as found in $LaSr_2Mn_2O_7$ and $La_{1/2}Ca_{1/2}MnO_3$. At any temperature we find no evidence of additional superlattice reflections with propagation vector ($h\pm1/4,k\pm1/4,0$) or ($h\pm1/2,k\pm1/2,0$) or ($h+e,0,l$), that may indicate such an orbital ordering. Recently Takata *et al.*[20] have shown that in the type-A antiferromagnet $NdSr_2Mn_2O_7$, charge density measurements are consistent with charge residing in Mn $d_{x^2-y^2}$ orbitals in the *ab*-plane, in the absence of any charge ordering, and a similar orbital ordering may occur in $LaSr_2Mn_2O_7$ at low temperatures.

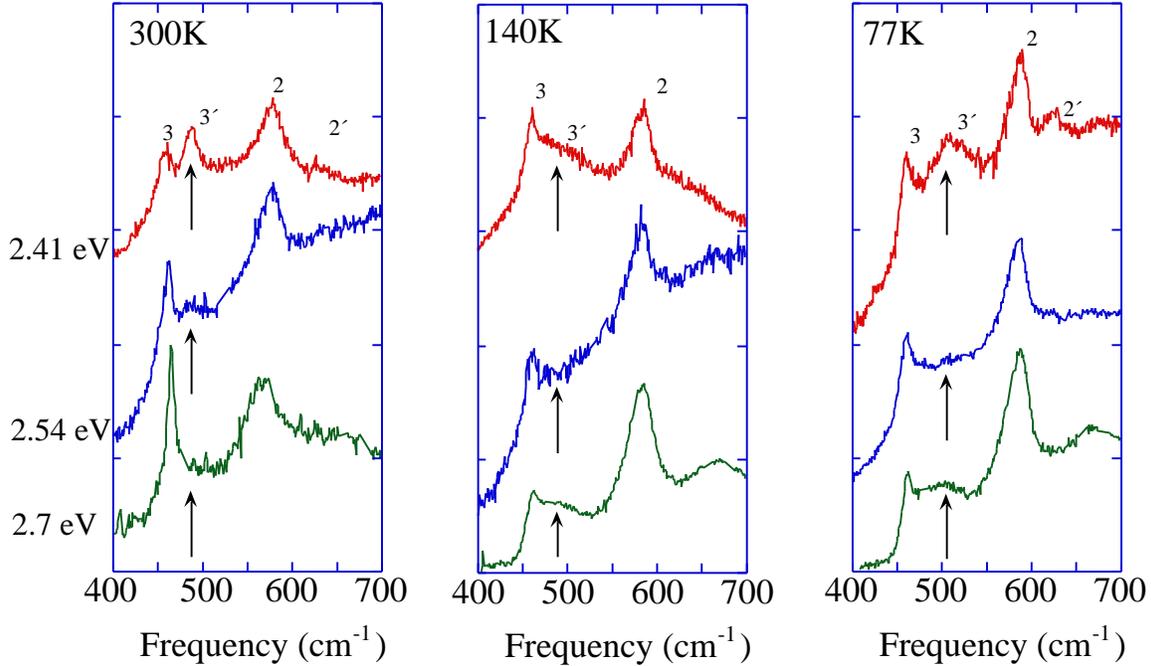

FIG. 4. Raman spectra measured at 300,140, and 77K using excitation wavelengths of 2.41, 2.54 and 2.7 eV.

## IV. LATTICE INSTABILITIES IN LaSr$_2$Mn$_2$O$_7$

Raman scattering involves the inelastic scattering of light by electrons and thus provides a tool to probe not only the lattice vibrations of a system but also variations in charge density.[21,22] The technique of resonant Raman scattering, where the light involved in the scattering process is resonant with a strong electronic transition or absorption edge is well-known and has been extensively used in semiconductors and high-T$_c$ materials.[23] In particular, if the energy of incident light is close to or in resonance with an electronic transition a dramatic increase in the Raman intensity may be observed, as reported for example in YBa$_2$Cu$_3$O$_7$ and La$_{1.67}$Sr$_{0.33}$NiO$_4$.[24,25]

Raman scattering spectra were measured using three different energies (2.41, 2.54 and 2.7eV), as a function of temperature from the powder sample of LaSr$_2$Mn$_2$O$_7$ (Fig. 4). At all excitation wavelengths the expected A$_{1g}$ Mn-O lattice vibrations are observed (see fig. 4) consistent with the single crystal Raman measurements of La$_{1.2}$Sr$_{1.8}$Mn$_2$O$_7$ by Romero $et.$ $al.$ [26] In particular we clearly observed the $in$-$plane$ Mn-O(3) bending mode ($\nu_3$) at 456 cm$^{-1}$ and the $apical$ Mn-O(2) stretching ($\nu_2$) mode at 575 cm$^{-1}$ [26] at 300K for all three excitation wavelengths (see fig. 3(a) and fig. 2 for labeling of atoms). Moreover, the spectrum taken at 2.41eV (514nm) clearly shows two peaks that exhibit a resonant enhancement at approximately 490 cm$^{-1}$ ($\nu_3'$), and 630 cm$^{-1}$ ($\nu_2'$) at 300K, as indicated by arrows on fig 4. Correlations in the change of the intensities of these new resonant Raman modes as a function of temperature and frequency provide insight into the lattice instabilities near the CO and AF regions. The resonant enhancement is consistent with a break in the symmetry of the electronic ground state of the system that gives rise to a resonance condition with an available excited $electronic$ state in LaSr$_2$Mn$_2$O$_7$ as has been observed in TTF-TCNQ and MX-linear chain complexes.[5]

In fig. 5 we show that the temperature dependent resonant Raman spectra measured using an incident laser energy of 2.41 eV. These spectra reflect both the effects of the charge ordering as seen in the resonant behavior but also the phase co-existence of the CO and AF phases as discussed in the previous section. As we shall show below, we find that changes in the in-plane modes are sensitive to the charge ordering transition while the apical modes are sensitive to the magnetic ordering and phase co-existence as discussed in the previous section. We shall deal with the two effects separately.

$In$-$plane$ $modes.$ — The temperature dependence of the frequency of the $\nu_3$ and $\nu_3'$ show an anomalous behavior (see fig. 6(a)). The $\nu_3$ mode softens by 3 cm$^{-1}$ between 340 and 240K, while both modes show an anomaly between 240 and 120K were their frequency is constant as a function of temperature. Below 120K both modes return to a classical behavior and harden. The similarities of the frequency of the $\nu_3$ and $\nu_3'$ modes and their respective temperature dependence suggest that they are coupled and of the same parentage. In addition the plateau in the temperature dependence of the frequency correlates with the observation of superlattice lines in the neutron and x-ray data and agrees with the symmetry of the $\nu_3$ and $\nu_3'$ modes as they arise from in-plane vibrations. Similar resonant behavior has been observed in the charge ordered nickelate $La_{1.67}Sr_{0.33}NiO_4$; resonant Raman modes appear at $T_{CO}$ resulting from the different in-plane Ni-O bonds. Similarly here the resonant Raman behavior of in-plane Mn-O(3) vibrations arises from the different bond lengths associated with the nominally $Mn^{3+}O_6$ and $Mn^{4+}O_6$ octahedra. Our observations differ compared to the $La_{1.67}Sr_{0.33}NiO_4$ compounds in that the $\nu_3'$ mode is observed in the 2.41 eV spectra as high as 340 K, and suggests that critical charge fluctuations are present prior to the CO transition at 210K. Associated with charge fluctuations a Jahn-Teller coupling to phonon modes is expected, as the crystal structure would be modulated locally around hopping $e_g$ charge carriers. With decreasing temperature our measurements suggest that charge-lattice fluctuations become static in agreement with the observed symmetry lowering at $T_{CO}$ and increase in resistivity as $T \to T_{CO}$. In addition Raman intensity is also observed in the 2.54 and 2.7 eV spectra were the $\nu_3'$ mode is expected (see fig. 3) indicating a lowering of the crystallographic symmetry.

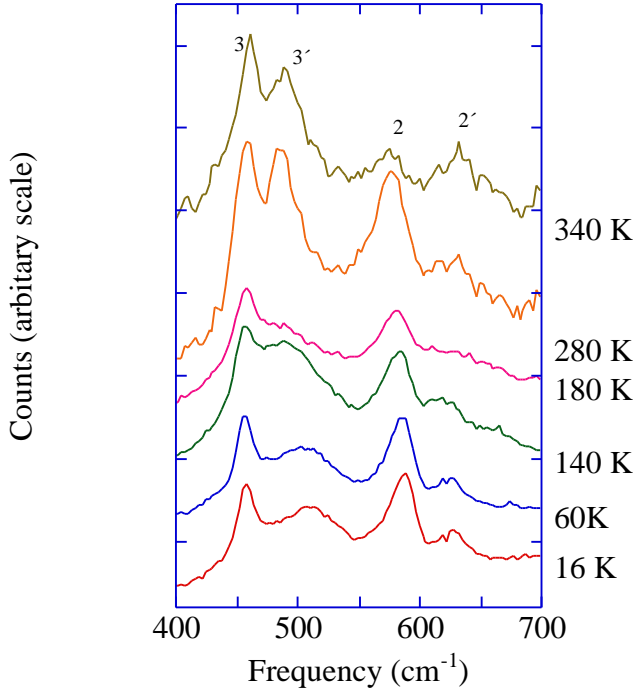

FIG. 5. Raman spectra as a function of temperature measured using an excitation wavelength of 2.41 eV.

An overall increase in the intensity of the in-plane modes with decreasing temperature is also observed as shown in fig 1(d). Here we plot the temperature dependence of the in-plane Mn-O(3) bending mode intensity, integrated from 440 to 545 cm$^{-1}$. The integrated intensity shows a strong enhancement in the charge ordered region, which reflects redistribution of the electronic spectra and tracks the resistivity of the material as well as the intensity of the observed superlattice reflections. We note here that although there is an overall enhancement in the integrated intensity of the in-plane Mn-O(3) modes the temperature dependence of the width of these modes is peculiar (see fig. 6(a)). The width of the $\nu_3$ mode shows an abrupt decrease at $T_{CO}$ consistent with the observed phase transition. However, the $\nu_3'$ mode shows completely the opposite behavior. Considering that the diffraction measurements clearly show a competition between the AF and CO phases, we suggest that this unusual broadening may reflect this competition. For example the pinning of CO domains around point defects at low temperatures.

*Apical Modes.* — As with the in-plane Mn-O(3) modes there is a smaller resonance effect also associated with the Mn-O(2) apical modes $\nu_2$ and $\nu_2'$. From a structural perspective, variations in the charge density within the *ab*-plane can result in different apical Mn-O bond for nominally $Mn^{3+}$ and $Mn^{4+}$ ions. These changes appear to be small, as they were not detected in our diffraction measurements, however they are apparent in the Raman spectra. For $\nu_2$ we find a classical behavior with the hardening of this mode with decreasing temperature (see fig. 6(b)). On the other hand frequency of the $\nu_2'$ mode decreases slowly with temperature till ~180 K and then precipitously decreases at $T_N$. Below $T_N$ this mode exhibits a classical behavior. In addition the width of $\nu_2'$ shows an increase with decreasing temperature and then a sharp decrease at ~$T_N$ while the width of the $\nu_2$ mode shows a classical behavior (see fig. 6(b)).

The changes in the Mn-O(2) stretching modes correlate with the development of the AF phase. As shown in the previous section, an unusual temperature dependence of the line width of the (0 0 10) reflection suggest that between $T_{CO}$ and $T^*$ there is a co-existence and competition between the CO and AF phases. In addition these data show that there is a pronounced difference between the c-axis lattice parameters of these two phases consistent with the pronounced difference observed in the apical Mn-O(2) modes.

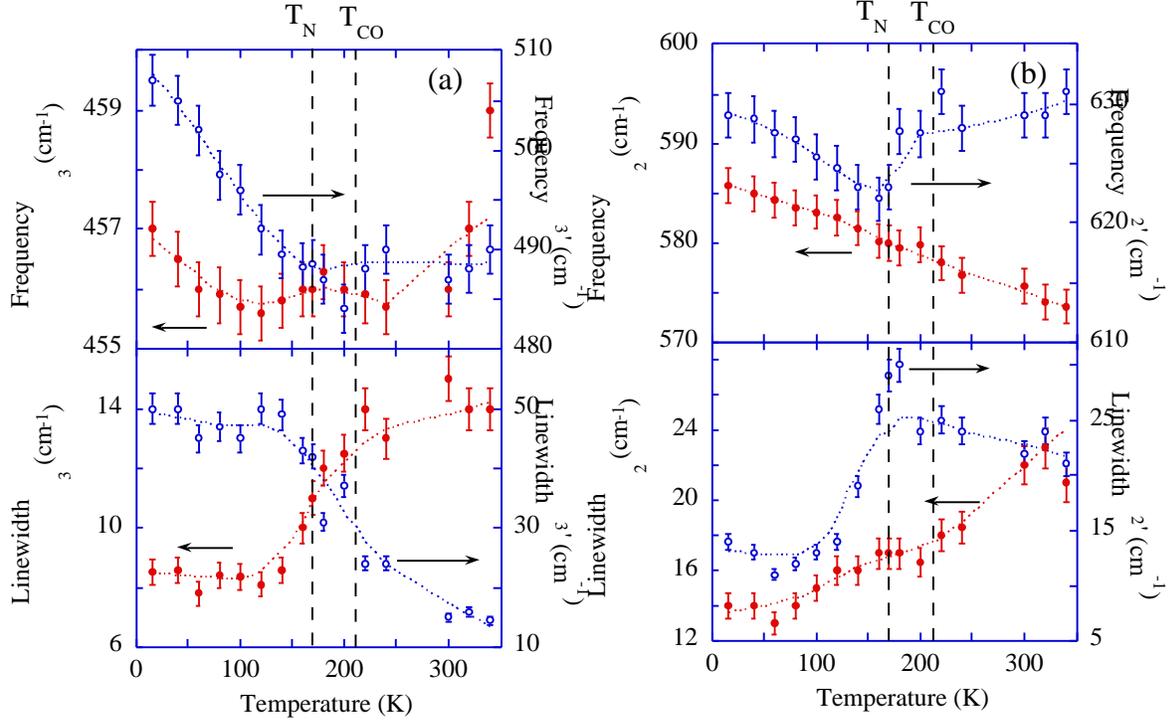

FIG. 6. Temperature dependence of the frequency and linewidths of (a) the $\nu_3$ and $\nu_3'$ and (b) $\nu_2$ and $\nu_2'$ Raman modes measured with an excitation wavelengths of 2.41 eV.

## V. DISCUSSION

Charge-lattice fluctuations (electron-phonon coupling), which arise from dynamic hopping mechanisms such as polarons, are now generally accepted to play an important role in the charge transport properties of the manganite perovskites. Here the Raman scattering provides an insightful picture for the region above $T_{CO}$. The resonant enhancement of modes coupled to *in-plane* vibrations suggest that dynamic charge-lattice fluctuations are present in $LaSr_2Mn_2O_7$ as high as 340K. Although we have not investigated the dispersion of these modes through the Brillouin zone, the observation of these fluctuations may be akin to dynamic charge ordering similar to what has been observed in $La_{1.85}Sr_{0.15}CuO_4$ by McQeeny *et al.*[1]. What is clear from our Raman and diffraction measurements is that these charge-lattice fluctuations are a precursor to the CO transition at $T_{CO}$.

The strong electron-phonon coupling associated with the charge ordered transition in $La_{1/2}Ca_{1/2}MnO_3$ has recently been investigated using infrared spectroscopy and has been described in terms of a charge density wave.[27] These measurements show that a BCS-like gap $2\Delta(T)$ fully opens at low temperatures and follows the hysteretic ferromagnetic-antiferromagnetic transition. Currently there are no such measurements for $LaSr_2Mn_2O_7$ to confirm weather the mechanism for the charge ordering transition is the same as in $La_{1/2}Ca_{1/2}MnO_3$. The measurements presented in this paper do not probe similar anomalies in the infrared spectrum or on the fermi surface, however, the observation of a coherent charge propagation vector, and resonant Raman behavior are consistent with a charge density wave picture.

In broader terms, this work provides insight into the charge-lattice dynamics of the layered manganites. Dynamic and/or static charge-lattice fluctuations may be a typically feature of these materials. They appear to be stable above magnetic transitions in layered CMR manganites and melt with the development of in-plane ferromagnetic interactions, resulting in a reduction of the resistivity. The correlation length of these charge-lattice correlations may vary from long range charge ordering as in $LaSr_2Mn_2O_7$ to shorter-range charge order correlations or polarons. Recent measurements by Doloc *et al.*[28] have shown a somewhat similar behavior in the ferromagnetic metallic layered manganite $La_{1.2}Sr_{1.8}Mn_2O_7$. In this material charge correlations are observed with propagation vector $\mathbf{q}=(h+\epsilon,0,l=2n+1)$ above $T_C$ and with coherence lengths of

~20 Å and melt close to the insulator-metal and ferromagnetic transitions.

## VII. CONCLUSION

The observation of charge-lattice correlations above $T_{CO}$, an accurate description of the crystallographic structure of the CO-phase and the competition between CO- and AF-phases constitute an important step in the understanding of the charge dynamics in the layered manganite perovskites. From our measurements it is clear that charge-lattice fluctuations above $T_{CO}$ become static with decreasing temperature and freeze to form a charge ordered phase. This freezing is manifested in both the observation of super lattice reflections below $T_{CO}$ and the increase in resistivity. Our analysis of single crystal neutron diffraction measurements shown that the orbital ordering associated with charge ordering is similar to that found in the perovskite manganites and predicted by Goodenough. However, in stark contrast to the perovskite manganites, the magnetic arrangement observed in the layered $LaSr_2Mn_2O_7$ is that of a type-A antiferromagnet that is incompatible with the observed CE-type orbital ordering. This incompatibility is resolved in that charge ordering and type-A antiferromagnetism co-exist as distinct and separate phases as shown by our X-ray diffraction measurements. With decreasing temperature the transition from charge ordering to type-A antiferromagnetism is characterized by a phase competition between these two.

## ACKNOWLEDGEMENTS

The authors thank J.D. Jorgensen for helpful discussions and comments on the manuscript. The work was supported by the U.S. Department of Energy, Basic Energy Sciences-Materials Sciences under contract W-7405-ENG-36 (DNA, HNB, JSG) and DE-AC02-98CH10886 (DEC), and UCDRD grant No. STB-UC:97-240 (HNB, GFS, KH) and by the MRL Program of the National Science Foundation under Award No. DMR96-32716 (BJC, AKC, ADS).